\documentclass[aip,pof,preprint]{revtex4-1}                  
             
\usepackage{graphicx}
\usepackage{amssymb}
\usepackage{amsmath}
\usepackage{subfigure,wrapfig}
\usepackage{color}

\begin{document}

\title{Vortex motion around a circular cylinder}

\author{G.~L.~Vasconcelos}\email[Corresponding author. Electronic mail: ]{giovani@df.ufpe.br}
\author{M.~N.~Moura  } 
\author{A.~M.~J.~Schakel}

\affiliation{Laborat\'orio de F\'{\i}sica Te\'orica e Computacional, 
Departamento de F\'{\i}sica, Universidade Federal de Pernambuco,
50670-901, Recife, Brazil.}

\date{November 18, 2011}

\begin{abstract}
  The motion of a pair of counter-rotating point vortices placed in a
  uniform flow around a circular cylinder forms a rich nonlinear system
  that is often used to model vortex shedding.  The phase portrait of
  the Hamiltonian governing the dynamics of a vortex pair that moves
  symmetrically with respect to the centerline---a case that can be
  realized experimentally by placing a splitter plate in the center
  plane---is presented.  The analysis provides new insights and reveals
  novel dynamical features of the system, such as a nilpotent saddle
  point at infinity whose homoclinic orbits define the region of
  nonlinear stability of the so-called F\"oppl equilibrium.  It is
  pointed out that a vortex pair properly placed downstream can overcome
  the cylinder and move off to infinity upstream.  In addition, the
  nonlinear dynamics resulting from antisymmetric perturbations of the
  F\"oppl equilibrium is studied and its relevance to vortex shedding
  discussed.
\end{abstract}

%\pacs{47.32.cb, 47.15.km, 47.27.wb}

\maketitle


\maketitle

\section{Introduction}
\label{intro}

Flow around a circular cylinder is a classical topic in hydrodynamics that
is of fundamental importance to many scientific fields with numerous
applications \cite{mmz, sf}. Of particular interest is the formation, at
moderate Reynolds numbers, of vortex eddies behind a circular cylinder,
which then go unstable at higher Reynolds numbers and evolve into a
Karman vortex street \cite{saffman,vvd}.  Since an
  analytic treatment of the problem in terms of the Navier-Stokes
equation is difficult  and the computational cost of direct numerical simulation very high, a particularly useful approach to study the basic
features of vortex shedding from bluff bodies is to consider the dynamics of 
point vortices in an inviscid fluid. 

A point-vortex model for the formation of two recirculating, symmetric
eddies in the wake of a circular cylinder was first introduced by
F\"oppl \cite{foeppl}.  He obtained stationary solutions for   a  pair 
  of vortices behind the cylinder in a uniform stream  and found that the centers of the eddies observed in the experiments lie on the locus of such equilibria---now called the F\"oppl curve. In addition,
  F\"oppl found that
these equlibria, although stable against perturbations that are symmetric
with respect to the centerline, were unstable against nonsymmetric
perturbations.  This instability is believed to constitute the origin of
the vortex shedding process that leads to the formation of the Karman
vortex street \cite{tang}.  It was later found out independently by
several authors \cite{smith,soibelman1,cai} that F\"oppl's stability
analysis for symmetric perturbations was in error in that the stationary solution  behind the cylinder
is not exponentially but only marginally stable.  Physically, marginal
stability implies, for instance, that if a splitter plate is placed
behind the cylinder in the center plane of the wake to suppress vortex
shedding \cite{roshko,roshko2,cai2003}, oscillating forces on the
cylinder may still arise owing to the cyclic motion of the vortices
around their equilibrium position \cite{laat}.
  
  Despite many contributions to the problem, it is fair to say that the
  nonlinear dynamics of the F\"oppl system is not yet fully
  understood. In particular, a more complete picture of 
  vortex-pair dynamics
  in the presence of symmetric perturbations is lacking,
  and several aspects of the nonlinear dynamics for nonsymmetric
  perturbations remain unclear.  To
  address these two issues is the main motivation of the present paper.
  It should be emphasized at the outset that a better understanding of
  the dynamical structure underlying the F\"oppl model is of interest
  not only  because of its practical relevance for vortex
  shedding, but also in its own theoretical right from the  viewpoint  of nonlinear dynamics.

  The F\"oppl model has inspired  a
    number of studies on several related problems, such as the modeling
  of vortex wake behind slender bodies in terms of multiple pairs of
  point vortices \cite{seath, weihs,miller,protas3}, the Hamiltonian
  structure of a circular  cylinder interacting dynamically with point vortices
  \cite{marsden,shashi2006,borisov2007}, the control of vortex shedding
  \cite{tang2000,protas1,protas2}, and the stability of symmetric and
  asymmetric vortex pairs over three-dimensional slender conical bodies
  \cite{cai2003, cai2005,bridges}. The related problem of
  desingularization of the  F\"oppl pair in terms of
  vortex patches of finite area was also  studied
  \cite{elcrat1,elcrat2}.  A recent review on vortex motion past
  solid bodies with additional references to the
  F\"oppl model and related problems can be found in
  Ref.~[28].

After formulating the problem of a
    pair of counter-rotating point vortices placed in a uniform stream
    around a circular cylinder in Sec.~\ref{sec:2}, we begin our
  analysis of the F\"oppl system  in Sec.~\ref{sec:3} by
  studying its Hamiltonian dynamics restricted to the invariant subspace
  where the vortices  move symmetrically
 with respect to the centerline.  A phase portrait of
  the system is presented that  fully
  characterizes the dynamics within this symmetric subspace.  In
    particular, we  point out
  that in addition to the two previously known sets of equilibira,
  namely, the F\"oppl equilibrium and the equilibrium on the axis
   bisecting the cylinder  perpendicularly to
  the uniform flow, the system possesses a hitherto unnoticed nilpotent saddle
  at infinity. We show furthermore that the homoclinic orbits associated with this nilpotent   saddle  delimit the region of closed orbits around the
  F\"oppl equilibrium.  We proceed in Sec.~\ref{sec:4} to study the
  linear and nonlinear dynamics resulting from antisymmetric
  perturbations of the F\"oppl equilibrium.  In the linear regime, a
  mistake that went undetected in F\"oppl's expressions \cite{foeppl}
  for the corresponding eigenvalues is now corrected.  As for the
  nonlinear dynamics, the unstable manifold associated with the F\"oppl
  equilibrium is computed numerically and its close relation to the
  vortex shedding instability is  pointed
    out.  The linear stability analysis of the equilibria on the normal
  line with respect to symmetric and antisymmetric perturbations is also
  presented---for the first time, it seems---and the respective
  nonlinear dynamics is investigated numerically. A discussion of the physical relevance of our findings and our main conclusions are presented in Sec.~\ref{sec:discuss}.

\section{Problem Formulation}
\label{sec:2}

\begin{figure}[t]
\begin{center}
\includegraphics[width=0.6\textwidth]{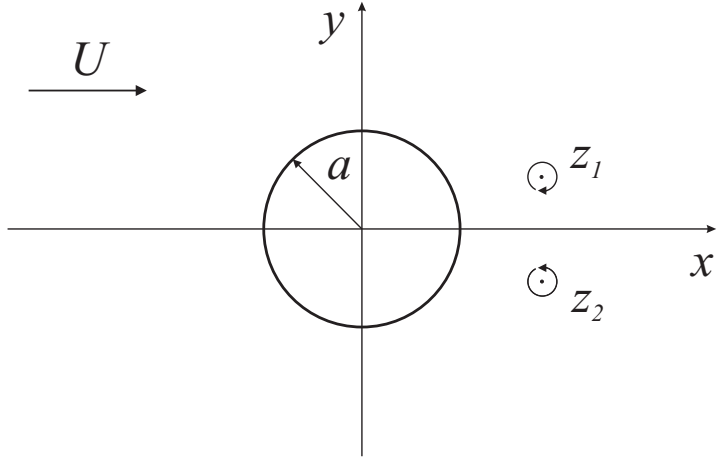}
\end{center}
\caption{A pair of vortices behind a circular cylinder in a uniform stream.}
\label{fig:1}
\end{figure}

We consider the motion of a pair of point vortices of same  strength
and opposite polarities around a circular cylinder of radius $a$ and in
the presence of a uniform stream of velocity $U$, as illustrated in
Fig.~\ref{fig:1}. It is convenient to work in the complex $z$-plane,
where $z=x+iy$, and place the center of the cylinder at the origin. The
upper and lower vortices are located at positions $z_1=x_1 + iy_1$ and
$z_2=x_2 + iy_2$, respectively. The complex potential
$w(z)=\phi(x,y)+i\psi(x,y)$, with $\phi$ being the velocity potential
and $\psi$ the stream function, is given by \cite{milne}
\begin{equation}
w(z)=U \left(z+\frac{a^2}{z}\right)+\frac{\Gamma }{2\pi  i}\ln 
\frac{z-z_1}{z- {a^2}/{\bar{z}_1}}-\frac{\Gamma }{2\pi  i}\ln \frac{z- z_2}{z-{a^2}/{\bar{z}_2}} ,
\label{eq:1}
\end{equation}
where $\Gamma$ is the circulation of the vortex at $z_{1}$ and bar
denotes complex conjugation. In Eq.~(\ref{eq:1}), the first two terms
represent the incoming flow and its image (a doublet at the origin) with
respect to the cylinder, the third term gives the contributions to the
complex potential from the upper vortex and its image, and similarly the
last term contains the contributions from the lower vortex and its
image. As can be inferred from Fig.~\ref{fig:1}, a necessary condition
for a steady configuration to exist is that the upper (lower) vortex be
of negative (positive) circulation, hence only the case $\Gamma<0$ is of
interest to us here.

In dimensionless variables
\begin{equation}
z'=\frac{z}{a}, \quad t'=\frac{U}{a}t, \quad w'=\frac{w}{Ua}, \quad \kappa=-\frac{\Gamma}{2\pi Ua}>0, 
\label{eq:non}
\end{equation}
the complex potential (\ref{eq:1}) becomes
\begin{equation}
w(z)= z+\frac{1}{z}+ i\kappa \ln
\frac{(z-{z_1})\left(1-\bar{z}_{2} z\right)}{(z- 
z_2)\left(1-\bar{z}_1z\right) } ,
\label{eq:12}
\end{equation}
where the prime notation has been dropped.  According to 
standard theory of point vortices in an inviscid fluid, any given vortex
moves with the velocity of the flow computed at the position of that
vortex, excluding its own contribution to the flow.  It then follows
from Eq.~(\ref{eq:12}) that the velocity  ${\bf u}_1=(u_1, v_1)$  of the vortex located at $z_1$ is
given by
\begin{equation}
u_{1}-iv_{1}=1-\frac{1}{z_{1}^2}-  i\kappa \left(\frac{1}{z_{1}-z_2}   - \frac{\bar{z}_1}{1-z_1\bar{z}_1}+ \frac{\bar{z}_2}{1-z_{1}\bar{z}_2}\right) ,
\label{eq:22}
\end{equation}
or more explicitly
\begin{subequations}
\label{eq:24}
\begin{align}
u_1=1 -\frac{x_1^2-y_1^2}{r_1^4}-  \kappa &  \left(\frac{y_1-y_2}{r_1^2+r_2^2-2(x_1x_2+y_1y_2)} + \frac{y_1}{r_1^2-1}- \frac{y_1r_2^2-y_2}{1+r_1^2r_2^2-2(x_1x_2+y_1y_2)} \right),
 \label{eq:24a}
\\
v_1= -2\frac{x_1y_1}{r_1^4} + \kappa & \left(\frac{x_1-x_2}{r_1^2+r_2^2-2(x_1x_2+y_1y_2)} + \frac{x_1}{r_1^2-1} -\frac{x_1r_2^2-x_2}{1+r_1^2r_2^2-2(x_1x_2+y_1y_2)}\right) ,
 \label{eq:24b}
\end{align}
\end{subequations}
where $r_i^2=x_i^2+y_i^2$, $i=1,2$.  The velocity ${\bf u}_2=(u_2, v_2)$  of the
second vortex is obtained by simply interchanging the indexes
$1\leftrightarrow 2$ in Eq.~(\ref{eq:24}) and letting
$\kappa\to-\kappa$.

\section{Dynamics on the Symmetric Subspace}
\label{sec:3}

It is not difficult to   see
  from Eq.~(\ref{eq:24}) that if the vortices are 
initially placed at positions symmetrically located  with respect to the
  centerline, i.e., $z_2(0)= \overline{z}_{1}(0)$, then this symmetry is preserved for all later times,
i.e., $z_2(t) =\overline{z}_1(t)$ for $t>0$.
In this section, we  study the dynamics within this invariant symmetric subspace, where the motion of the lower vortex is simply the mirror
  image of that of the upper vortex with respect to the centerline.
Symmetry can be enforced experimentally
by placing a splitter plate behind the cylinder in the center plane of the wake \cite{foeppl,roshko}. 

With $x_2=x_1$ and $y_2=-y_1$,  Eq.~(\ref{eq:24})  reduces to
\begin{subequations}
\label{eq:4}
\begin{equation}
u=1-\frac{x^2-y^2}{r^4}+  \kappa  y \left[\frac{r^2+1}{(r^2-1)^2+4y^2} -\frac{1}{r^2-1}
 - \frac{1}{2y^2}\right],
 \label{eq:4a}
\end{equation}
 \begin{equation}
v=-2\frac{xy}{r^4} - \kappa x \left[\frac{r^2-1}{(r^2-1)^2+4y^2}-\frac{1}{r^2-1}\right].
 \label{eq:4b}
\end{equation}
\end{subequations}
Here, the subscripts have been dropped with the understanding that in
the remainder of the section we restrict our attention to the upper
vortex. 

\subsection{Hamiltonian dynamics and phase portrait}

As is well known, the equations of motion for point vortices in a two-dimensional
inviscid flow, first derived by Kirchhoff, can be formulated as a Hamiltonian system \cite{saffman, vvd}.  The dynamics of point vortices in the presence of  closed, rigid boundaries   
was shown by Lin \cite{lin1941} to be also Hamiltonian with the same canonical sympletic structure as in the absence of boundaries.      For a vortex pair  placed  in a uniform stream around a circular cylinder, the phase space is four-dimensional 
and has a
two-dimensional (2D) invariant subspace corresponding to symmetric
orbits. 
The Hamiltonian  restricted to  the
2D symmetric subspace is given by \cite{zannetti}
\begin{equation}
H(x,y)=y\left(1-\frac{1}{r^2}\right)-\frac{\kappa}{2}\ln\frac{y(r^2-1)}{\sqrt{(r^2-1)^2+4y^2}}.
\label{eq:H}
\end{equation}
The corresponding dynamical equations
\begin{equation}
\label{eq:HJ}
\dot{x}=\frac{\partial H}{\partial y}, \quad \dot{y}=-\frac{\partial H}{\partial x},
\end{equation}
where dot denotes time derivative,  yield Eq.~(\ref{eq:4})
  upon identifying $(u, v)$ with $(\dot{x}, 
    \dot{y})$.
    
\begin{figure}
\begin{center}
\includegraphics[width=0.6\textwidth]{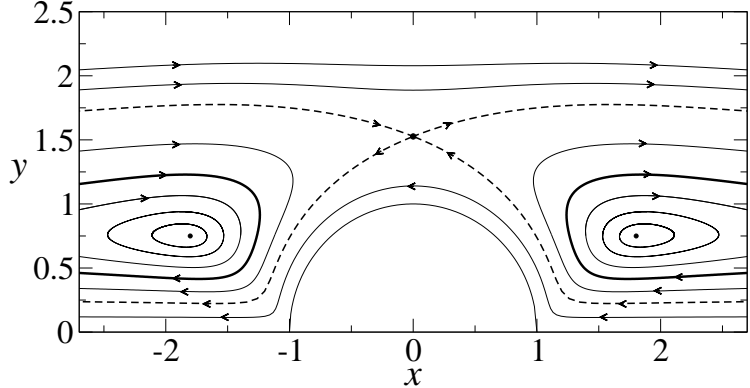}
\end{center}
\caption{Phase portrait for the symmetric   F\"oppl system with $\kappa=45/32$. The isolated black dots are the F\"oppl equilibria. The dashed curves are the stable and unstable branches of the separatrix associated with the equilibrium point on the normal line, and the thick solid lines are the homoclinic loops of the equilibrium point at infinity; see text.}
\label{fig:phase}
\end{figure}

A phase portrait of this 
  Hamiltonian system for $\kappa=45/32$ is presented in Fig.~\ref{fig:phase},  where the curves shown are (unevenly spaced)  level sets of the Hamiltonian (\ref{eq:H}). [For convenience, these curves were obtained from a direct numerical integration of Eq.~(\ref{eq:4}).] A
detailed description of the  main features of this
 phase portrait will be given below,
  starting with an analysis of the various equilibrium points and their
  stability.
The related problem of the symmetric ``moving F\"oppl system,''  where the cylinder advances through the fluid  followed  by the vortex pair, was recently considered by Shashikanth {\it et al.} \cite{marsden}, but there the phase portrait \cite{shashi2006} is quite different from the one shown in Fig.~\ref{fig:phase}, because of  the additional degrees of freedom related to the velocity of the moving cylinder. 

\subsection{Equlibrium points}

The equilibrium positions for the vortex are obtained by setting $u=v=0$ in Eq.~(\ref{eq:4}). Three types of equilibrium points can be identified.

\subsubsection{F\"oppl equilibria} 
The
locus of possible equilibrium positions $(x_0,y_0)$ for the upper vortex  found by F\"oppl \cite{foeppl}
is the curve
\begin{equation}
r_0^2-1=  2 r_0 y_0,
\label{eq:pair}
\end{equation}
with corresponding strength 
\begin{equation}
\kappa =\frac{(r_0^2+1)(r_0^2-1)^2}{r_0^5}.
\label{eq:7}
\end{equation}
Along the  F\"oppl curve (\ref{eq:pair}), the vortex strength increases with distance from the center
of the cylinder and diverges linearly for $r_0 \to \infty$.  For the
equilibrium point on the edge of the cylinder ($r_0 \to 1$), the
strength vanishes.
Notice that Eq.~(\ref{eq:pair}) yields two branches of solution: one in
which the vortex pair is behind the cylinder ($x_0>0$) and the other
where the vortex pair is in front of the cylinder ($x_0<0$). The former
case models the formation of vortex eddies behind a cylinder in a
uniform stream and was the primary motivation of F\"oppl's original study
\cite{foeppl}. The latter case  has attracted far less attention because it is not usually observed in experiments. We note, however, that recirculating eddies are observed in front of a circular cylinder near a plane boundary when the gap between the cylinder and the plane is sufficiently small \cite{lin}.
In this context, the F\"oppl equlibrium upstream of the cylinder may eventually be relevant for flows around a
half-cylinder placed on a plane wall (or for the closely related
situation where a splitter plate is attached to the front of the cylinder), although we are unaware of specific experiments in this setting.
 
\subsubsection{Equilibria on the normal line}

This corresponds to the upper vortex being located on the line bisecting the cylinder perpendicularly to
the incoming flow \cite{weihs}, that is,
\begin{equation}
x=0, \qquad y= b, \qquad  b>1,
\label{eq:b}
\end{equation} 
with  strength
\begin{equation}
\kappa =\frac{2(b^2-1)(b^2+1)^2}{b(b^4+4b^2-1)}.
\label{eq:kappa}
\end{equation}
As in the F\"oppl solution, the strength tends to zero when the
  edge of the cylinder is reached ($b \to 1$) and diverges linearly with
  distance from the center of the cylinder.  At large distances, the
  vortex strength for this equilibrium is about twice that of a F\"oppl
  pair located at the same distance from the origin.

\subsubsection{Equilibrium at infinity}

Equation~(\ref{eq:4})  also  yields  equilibrium points at the  positions
\begin{equation}
x=\pm\infty, \qquad y_\infty=\frac{\kappa}{2} .
\label{eq:yc}
\end{equation}
To the best of our knowledge, the existence of this additional
equilibrium point at infinity was not noted before. Its physical origin,
however, can be easily understood, as it corresponds to the equilibrium
configuration for a vortex pair placed in a uniform stream (without the
cylinder).  At points infinitely far from the cylinder, the flow induced
by the image system (inside the cylinder) becomes
negligible and hence a stationary configuration is possible if the
vortices with given circulation $\pm \kappa$ are placed
at the appropriate distance ($=\kappa$) from each other.

\subsection{Stability analysis}
\label{sec:sasym}

The linear stability analysis of the equilibria described above is presented next,   together with a discussion of the  \emph{nonlinear} stability of the F\"oppl equilibrium. 

\subsubsection{F\"oppl equilibria} 

Consider a
perturbation of the F\"oppl equilibrium (\ref{eq:pair})
 parameterized as: $z=z_0+{\Delta z}$, where ${\Delta z} =
\xi+i\eta$, with $\xi$ and $\eta$ being infinitesimal (real)
quantities. Linearization of Eq.~(\ref{eq:4}) then yields the following
dynamical system
\begin{equation}
 \left(\begin{array}{c} \dot{\xi}\cr \dot{\eta}\end{array}\right)= 
A  \left(\begin{array}{c}{\xi}\cr{\eta}\end{array}\right),
\end{equation}
where the matrix $A$ reads
\begin{equation}
 A_{11} =-A_{22}=-\frac{x_0(r_0^4 - 3 r_0^2+2)}{r_0^8},
\end{equation}
\begin{equation}
 A_{12} =\frac{
 4r_0^8+5r_0^6+2r_0^4-5r_0^2+2}{2r_0^9},
 \end{equation}
\begin{equation}
 A_{21}  = -\frac{2x_0^2(r_0^4+r_0^2+2)}{r_0^7(r_0^2+1)}.
\end{equation}
 Its eigenvalues $\lambda$ are given by
\begin{equation}
\lambda^2 =- \frac{3 r_0^6+5 r_0^4+13 r_0^2-5}{r_0^{10}}<0,
\label{eq:ls}
\end{equation}
for $r_0>1$.  The eigenvalues are thus purely imaginary, and not a
complex pair with negative real part as found by F\"oppl \cite{foeppl}.
In other words, the F\"oppl equilibrium is a center and not a stable
focus. Our equation (\ref{eq:ls}) agrees with the
expression for the eigenvalues of the symmetric modes obtained in
Ref.~[7] from the linearization of the full 4D dynamical system.
 As can be seen from Fig.~\ref{fig:phase}, the F\"oppl solution
is in fact a nonlinearly stable center, meaning that when the vortex is
displaced from its equilibrium position by a small (but finite) amount,
it executes a periodic motion around that point, corresponding to the
closed orbits in the figure. 
 This periodic motion around the F\"oppl equilibrium has been
  observed in numerical simulations of the model carried out by de Laat
  and Coene \cite{laat}. Note that since the eigenvalues given in
Eq.~(\ref{eq:ls}) do not depend explicitly on the coordinate $x_0$, it
follows that the two F\"oppl equilibria, downstream and upstream of the
cylinder, have identical stability properties, as is evident from
Fig.~\ref{fig:phase}.  This means, in particular, that if vorticity can be generated upstream of the cylinder then stationary recirculating eddies could form in front of the cylinder---a situation observed, for instance, in flows around a
cylinder placed above a plane wall \cite{lin}.

\subsubsection{Equilibria on the normal line}
\label{sec:C2}

Linearization of Eq.~(\ref{eq:4}) around the equilibrium point $z=ib$
yields  for the matrix
$A$:
\begin{equation}
 A_{11} =A_{22}=0,
\end{equation}
\begin{equation}
 A_{12} =\frac{b^8+10 b^6-8 b^4+14 b^2-1}{b^3(b^2-1)(b^4+ 4 b^2-1 )},
 \end{equation}
\begin{equation}
 A_{21}  = \frac{2(b^2-1)(3b^2-1)}{b^3(b^4+ 4 b^2-1 )}.
\end{equation}
The eigenvalues $\lambda$  of this
  matrix are determined by
\begin{equation}
\lambda^2=\frac{2 \left(3 b^2-1\right) \left(b^8+10 b^6-8 b^4+14 b^2-1\right)}{b^6 (b^4+ 4 b^2-1 )^2}>0,
\end{equation}
 which yields a pair of real
eigenvalues, $\lambda_\pm=\pm\sqrt{\lambda^2}$. The equilibrium point on the
normal line is therefore a saddle,  having a stable and unstable direction,  as is  also evident from the phase
portrait shown in Fig.~\ref{fig:phase}.  The eigenvectors ${\bf w}_\pm$
associated with the eigenvalues $\lambda_\pm$, respectively, read
\begin{equation}
  {\bf w}_\pm= \left(\begin{array}{c}   \pm \sqrt{ A_{12}/A_{21}} \cr 1\end{array}\right). 
\label{eq:theta}
\end{equation}
Although it was known from numerical simulations \cite{laat} that the
equilibrium point on the normal line is unstable (against generic symmetric perturbations), it seems that an
explicit linear stability analysis for this case was not carried out
before, perhaps because these equilibria were not considered physically
relevant since they are not observed in
experiment \cite{foeppl}. 
However, when the full nonlinear dynamics is considered,
the stable and unstable eigendirections ${\bf w}_\pm$   give origin to the respective stable and
unstable separatrices, indicated by the dashed curves in
Fig.~\ref{fig:phase}. In this sense,  the existence of an equilibrium point on the normal line is dynamically felt by a vortex even if it is placed far from this ``unphysical'' equilibrium.  

\subsubsection{Equilibrium at infinity}

\begin{figure}
\begin{center}
\includegraphics[width=0.6\textwidth]{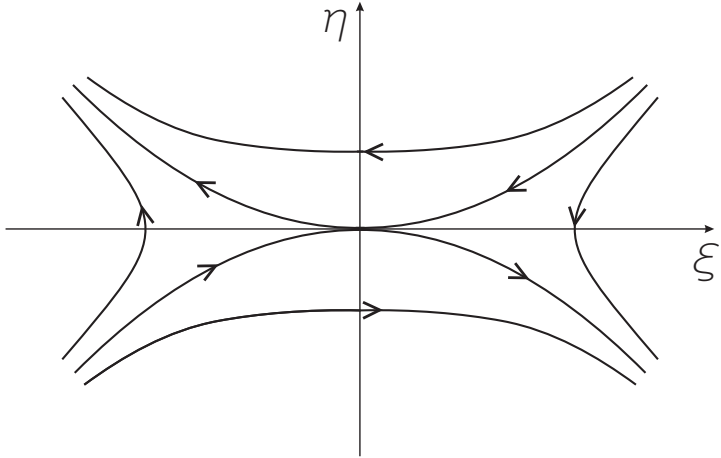}
\end{center}
\caption{Trajectories near a nilpotent saddle.}
\label{fig:saddle}
\end{figure}

 The matrix $A$ of the  linearized system 
 around the equilibrium point at infinity is   given by
\begin{equation}
 A=  -\frac{2}{\kappa} \left(\begin{array}{cc}    0 & 1 \cr 0 & 0\end{array}\right),
\end{equation}
 which is nilpotent and has  two zero eigenvalues.    
To study the stability of this equilibrium point,  one needs to examine the nonlinear contributions. To this end, we note that for
$|x| \to \infty$ and $y \approx y_\infty$, Eq.~(\ref{eq:4b}) assumes the
form
\begin{equation}   
\label{eq:doty}
\dot{y} = -\frac{\kappa}{x^3} .
\end{equation} 
It then follows from a theorem in ordinary differential equations
\cite{perko} that, in view of the cubic term in Eq.~(\ref{eq:doty}), the
equilibrium point  is a  \emph{degenerate} or \emph{nilpotent} saddle \cite{abraham}, for  which the two
eigenvectors are the same. The behavior of trajectories in the neighborhood of a generic nilpotent saddle is illustrated in Fig.~\ref{fig:saddle}. The behavior
near the nilpotent saddle at $x=\pm\infty$ and $y=y_\infty$ can be
described as follows.  A vortex placed very far downstream and
below (above) the line $y=y_\infty$ will move away from
(towards) the equilibrium point at $x=\infty$. Similarly, a vortex
placed very far upstream  will move away from (towards) the
equilibrium point at $x=-\infty$ if $y>y_\infty$
($y<y_\infty$). 

The stable and unstable separatrices 
associated with the nilpotent saddle at infinity form two
homoclinic loops \cite{abraham},  called nilpotent saddle loops, which  are indicated  in Fig.~\ref{fig:phase} by 
thick solid lines and  correspond to the level curves passing through  this equilibrium point: 
\begin{equation}
  H(x,y)=H(\pm\infty,y_\infty)=\frac{\kappa}{2} \left(1- \ln \frac{\kappa}{2} \right).
\end{equation}
The nilpotent saddle loops 
 encircle the F\"oppl equilibria and define
their region of nonlinear stability, in   
the sense that vortex trajectories are closed for initial positions  inside the loops and unbounded otherwise.   In this way, the nilpotent saddle at
  infinity, which went unnoticed until now, allows us to  fully characterize the nonlinear stability of the F\"oppl equilibrium.

For unbounded orbits, the long-time asymptotic behavior depends on the
location of the vortex initial position with respect to the separatrices
associated with the equilibrium point on the normal line.  A vortex
placed downstream of the cylinder between the nilpotent saddle loop and the separatrices of the equilibrium point on normal line will eventually be
convected away by the free stream; see Fig.~\ref{fig:phase}. In
particular, if the vortex starts very far behind the cylinder at a
position that is below the nilpotent saddle loop and above the stable
separatrix, it first moves towards the cylinder, turns around the
F\"oppl equilibrium, and is then ``reflected'' back to infinity.  Even
more surprising trajectories arise if the vortex is placed downstream
below the stable separatrix, for it will be close enough to its image
below the centerline to be able to overcome the cylinder and move off to
infinity upstream.  (A related phenomenon occurs in the
  inviscid coupled motion of a cylinder initially at rest and a vortex pair starting at infinity with no
  imposed background flow \cite{eames}. When the cylinder is less dense than the fluid, it is found that if the vortices are released
  sufficiently above the centerline  they reverse 
  relative to the moving cylinder; otherwise, they move over and past
  the cylinder.)  Unbounded trajectories
for the F\"oppl system 
  also result for initial
positions upstream of the cylinder: i) if placed above the stable separatrix, the vortex moves downstream to infinity;  and ii) if placed between the stable separatrix and the nilpotent saddle loop, the vortex goes  around the F\"oppl equlibrium in front of the cylinder and returns to infinity upstream; see Fig.~\ref{fig:phase}. It is again
the hitherto unnoticed nilpotent saddle at infinity, together with the
precise nature of the equilibrium point on the normal line, that allows
us to go beyond linear stability analysis and capture the full phase
portrait in the symmetric subspace.

We stress that
closed orbits exist only when the flow is symmetric.  Nonsymmetric
 perturbations inevitably cause
the vortex pair to move off to infinity, as we  
 demonstrate  next.

\section{Nonsymmetric Dynamics}
\label{sec:4}

In this section, the effect of antisymmetric perturbations on the equilibria of the F\"oppl system is studied.  We begin
 by  observing that
the dynamics of two counter-rotating point vortices possesses a  \emph{conjugation symmetry}.  To describe this symmetry,
let  $z_1(t;z_{1,0},z_{2,0})$ and
$z_2(t;z_{1,0},z_{2,0})$  denote the trajectories of the
 upper and  lower
vortices, respectively, with
initial positions $z_{1,0}$ and $z_{2,0}$.  For the dynamical system  
  defined by Eq.~(\ref{eq:24})  and the corresponding equation for the second vortex,  
one can verify that the following relations hold
\begin{subequations} 
\label{eq:cc}
\begin{equation}
z_1(t;\overline{z}_{2,0},\overline{z}_{1,0})=\overline{z_2(t;z_{1,0},z_{2,0})} ,
\end{equation}
\begin{equation}
z_2(t;\overline{z}_{2,0},\overline{z}_{1,0})=\overline{z_1(t;z_{1,0},z_{2,0})} .
\end{equation}
\end{subequations}
%where we recall that bar denotes complex conjugation. 
In other words, for any given pair of initial positions,
$z_{1,0}$ and $z_{2,0}$, there 
exists a ``conjugate pair'' of initial positions,
$\overline{z}_{2,0}$ and
$\overline{z}_{1,0}$, such that the vortex trajectories
 of the first pair are the complex
conjugate of
 those of the second pair. 

Any perturbation of a vortex-pair equilibrium can be written as the superposition of a symmetric perturbation and an antisymmetric one. To be precise, antisymmetric perturbations are of the form
\begin{equation} 
z_1=z_0+\Delta z, \qquad z_2=\overline{z}_0-\overline{\Delta z},
\label{eq:daz}
\end{equation}
 where $z_0$ denotes a generic equilibrium point and ${\Delta z} = \xi+i\eta$.   Since the antisymmetric subspace of the full 4D phase space is invariant under linear dynamics, we can focus on the upper vortex in carrying out our linear stability analysis.

\subsection{F\"oppl equilibria}

\label{sec:unst}

 Linearization of Eq.~(\ref{eq:24}) around the F\"oppl equilibrium (\ref{eq:pair})  with respect to antisymmetric perturbations  (\ref{eq:daz}) yields
\begin{equation}
 \left(\begin{array}{c} \dot{\xi}\cr \dot{\eta}\end{array}\right)= 
B  \left(\begin{array}{c}{\xi}\cr{\eta}\end{array}\right) ,
\end{equation}
where the matrix $B$ is given by
 \begin{equation}
 B_{11}=-B_{22}=\frac{x_0\left(r_0^4+3 r_0^2-2\right)}{r_0^8},
 \label{eq:B11}
\end{equation}
\begin{equation}
B_{12} = \frac{3 r_0^6-5 r_0^2+2}{2 r_0^9},
\end{equation}
\begin{equation}
 B_{21}  = \frac{4 r_0^8+3 r_0^6 -4 r_0^4-5 r_0^2+2}{2 r_0^9}.
 \label{eq:B21}
\end{equation}
This matrix has a pair of real eigenvalues, $\lambda_{\pm}=\pm\sqrt{\lambda^2}$, where 
\begin{equation}
\lambda^2 =\frac{3 r_0^6+3 r_0^4-3 r_0^2 + 1}{r_0^{10}}.
\label{eq:las}
\end{equation}
The F\"oppl equilibrium is therefore a saddle with respect to
antisymmetric perturbations, while it is a center with respect to
symmetric perturbations, as seen earlier. That is, the F\"oppl
equilibrium is a {\it saddle-center} of the full 4D dynamical system
\cite{hs}. We note in passing that, although F\"oppl
 obtained a pair of real eigenvalues for the case of
antisymmetric perturbations, his original formulae for the eigenvalues
are in error \cite{footnote}.
Our expression (\ref{eq:las}) is in agreement with the eigenvalues of
the skew-symmetric modes obtained by Smith \cite{smith} from the
linearization of the full dynamical system.  The eigenvectors ${\bf
  w}_{\pm}$ associated with the eigenvalues $\lambda_\pm$ are readily
computed, with the result
\begin{equation}
{\bf w}_\pm = \left(\begin{array}{c} (\lambda_\pm+B_{11})/B_{21}  \cr 1\end{array}\right) .
\label{eq:waf}
\end{equation}

\begin{figure}[t]
\begin{center}
\includegraphics[width=0.6\textwidth]{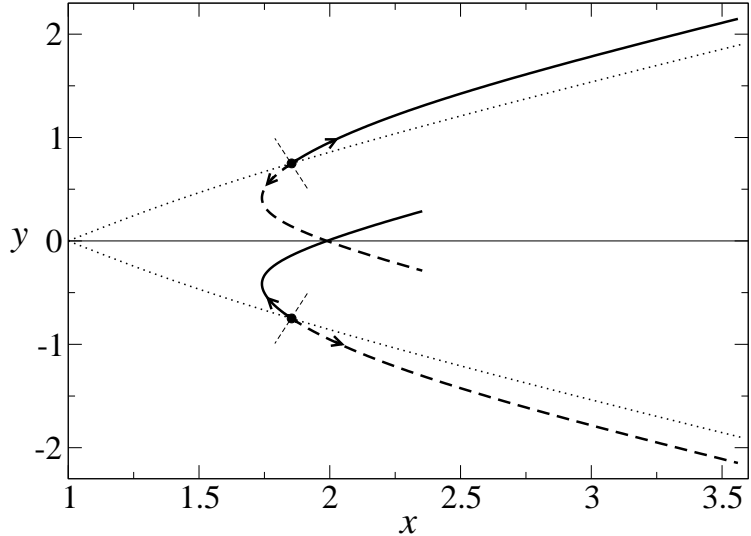}
\end{center}
\caption{Vortex trajectories for antisymmetric perturbations of the
  F\"oppl equilibrium for $\kappa=45/32$, in which case
  $x_0=\sqrt{55}/4$ and $y_0=\pm3/4$ (black dots).   The solid  and dashed curves are the trajectories starting along the unstable
 directions ${\bf w}_+$ and $-{\bf w}_+$,
    respectively, while the short straight lines  
  indicate the axes defined by the stable direction ${\bf w}_-$. The dotted lines represent the loci  of the F\"oppl equilibria.}
\label{fig:anti}
\end{figure}

 In Fig.~\ref{fig:anti}, we show in solid
 curves the  pair of vortex trajectories    obtained by slightly  displacing the vortices from  their equilibrium positions in the directions defined by  the unstable
eigenvector ${\bf w_+}$,    while the trajectories
  obtained by slightly displacing the vortices  in the opposite directions
 are shown in dashed curves.
  The latter pair of trajectories is the complex
conjugate of the former by conjugation symmetry.  Note that for the first pair  of trajectories, the lower
vortex initially moves towards the centerline and upstream, while the
upper vortex moves away from the centerline and downstream.  
 At later times, the vortex pair moves off to infinity
with the lower vortex trailing behind the upper vortex.  For 
 the second pair of trajectories, the upper and lower vortices switch
role; see Fig.~\ref{fig:anti}.   In the flow of a real fluid past a cylinder, the two
basic instabilities associated with  displacements along   the unstable directions $\pm{\bf w_+}$ happen alternately  and
constitute the origin of    vortex shedding    that
leads to  the  formation of the Karman vortex street
\cite{tang}. 
In like manner, the suppression of vortex shedding by placing  a splitter plate behind the cylinder \cite{roshko,roshko2} is
consistent with the fact that the F\"oppl equlibrium is nonlinearly
stable with respect to symmetric perturbations;   see Sec.~\ref{sec:discuss} for further discussions on vortex shedding and its suppression by a splitter plate.

For small, generic antisymmetric perturbations, the vortices move along
trajectories that follow closely the ones depicted in
Fig.~\ref{fig:anti}.  Whether a vortex pair eventually moves up or down
is determined by the initial position of the upper vortex relative to
the stable direction ${\bf w}_-$, which is indicated in
Fig.~\ref{fig:anti} by the short straight line passing through the
F\"oppl equilibrium. If the initial position of the upper vortex is to
the right (left) of the stable direction, then the vortex pair
asymptotically moves upwards (downwards).  
This explains the
behavior seen in the numerical
simulations reported in Ref.~[25], where nearby initial positions around
the F\"oppl equilibrium were found to lead to close-by
trajectories. 

\begin{figure}
\begin{center}
\vspace{0.5cm}
\includegraphics[width=0.6\textwidth]{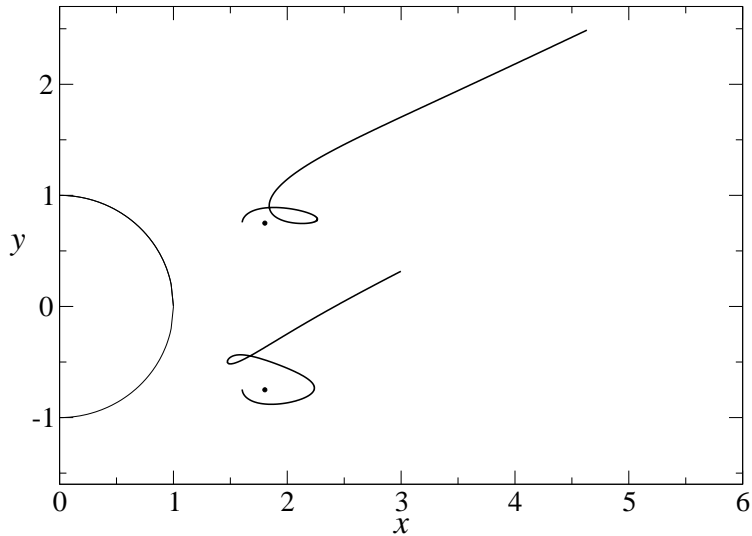}
\end{center}
\caption{Trajectories resulting from a generic perturbation ${\Delta z}_1=\Delta z_2=-0.25 + i0.005$ of the F\"oppl  pair at $r_0=2$ (black dots).}
\label{fig:loop}
\end{figure}

Since any degree of antisymmetry in the initial perturbation causes the
vortex pair to move off to infinity, the F\"oppl equilibrium is unstable
under generic perturbations.  As an example, Fig.~\ref{fig:loop} shows
vortex trajectories obtained by displacing the F\"oppl pair (at $r_0=2$)
by the amounts $\Delta z_1=\Delta z_2=-0.25+i0.005$. During the linear
stage, the trajectories are a superposition of a symmetric orbit and a
growing mode associated with the antisymmetric component of the
perturbation, which ultimately leads to asymptotic trajectories with the
vortices moving parallel to each other.  

\subsection{Equilibria on the normal line}

\label{sec:4b}

For antisymmetric perturbations of the equilibrium  (\ref{eq:b}) on the normal line, the
matrix $B$  assumes the form
 \begin{equation}
 B_{11}=B_{22}=0,
\end{equation}
\begin{equation}
B_{12}=\frac{2 \left(3 b^6+b^4+5 b^2-1\right)}{b^3(b^2-1)\left(b^4+4b^2-1\right)},
\end{equation}
\begin{equation}
  B_{21} = \frac{b^2-1}{b^3},
\end{equation}
with eigenvalues  $\lambda$ given by
\begin{equation}
  \lambda^2 =\frac{2 \left(3 b^6+b^4+5 b^2-1\right)}{b^6 \left(b^4+ 4 b^2-1 \right)}>0.
\end{equation}
This yields a pair of real eigenvalues,
$\lambda_{\pm}=\pm\sqrt{\lambda^2}$, with respective eigenvectors:
\begin{equation}
  {\bf w}_\pm = \left(\begin{array}{c} \pm \sqrt{B_{12}/B_{21}} \cr 1\end{array}\right) .
\label{eq:theta}
\end{equation}

\begin{figure}
\begin{center}
\includegraphics[width=0.6\textwidth]{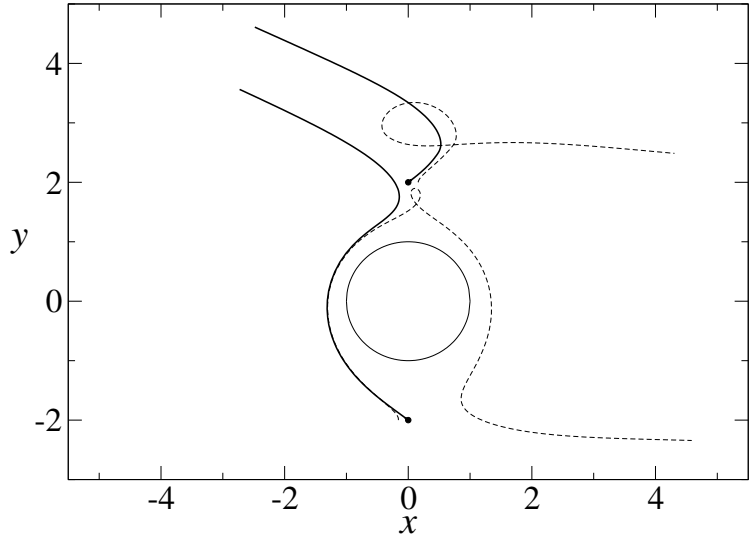}
\end{center}
\caption{Vortex trajectories (solid curves) associated with the unstable
  direction ${\bf w}_+$  of the equilibrium at
  $z=\pm2i$ (black dots). The dashed curves are trajectories resulting from the antisymmetric
  perturbation $\Delta z= 0.16$.}
\label{fig:normal}
\end{figure}

In Fig.~\ref{fig:normal}, we show the vortex trajectories (solid curves)
obtained by slightly displacing the vortices from their equilibrium
position along the unstable direction ${\bf w}_+$ for $b=2$.  The
initial motion here is somewhat similar to what is seen for a F\"oppl
pair, in the sense that one vortex moves upstream towards the centerline
and the other moves downstream away from the centerline.  The main
difference is that for later times, the vortices now end up moving
upstream. The long-time dynamics in this case is also more sensitive on
the initial conditions: for somewhat larger perturbations, the vortices
are eventually carried away by the free stream.  An example where this
happens is indicated by the dashed curves in Fig.~\ref{fig:normal},
which represent the vortex trajectories for the antisymmetric
perturbation $\Delta z=0.16$. 

 As already argued in Sec.~\ref{sec:C2},   although
 the equilibrium point on the normal line is not directly observed in experiments,
 it is important  to  know
   its  instability properties under both
  symmetric and antisymmetric perturbations.
This knowledge  contributes to a better understanding not only of the full nonlinear dynamics  of the F\"oppl system but also  of  more general flows,  such as the case of  stationary  vortex patches above and below the cylinder in a uniform stream, where similar unstable modes are observed \cite{elcrat2}.

\section{Discussion and Conclusions}
\label{sec:discuss}

In this paper, we have investigated a
 two-dimensional vortex model for the formation of
recirculating eddies behind a fixed cylinder placed on a uniform stream.   The model, which was first introduced by F\"oppl \cite{foeppl} almost a century ago, has two main
  simplifying assumptions: i) the fluid is treated as inviscid and hence
  the flow is potential, and ii) the size of the vortex core is
  neglected and so the vortices are considered to be point-like.  In
spite of  these simplifications, the
model is known to be in qualitative
 agreement with real flows past a
cylinder, as  was already
   pointed out by F\"oppl in his original paper. Several novel features of the F\"oppl model have been obtained in the present work, which help one to better understand the basic dynamics of vortex shedding behind a cylinder.

In real flows,  governed by the  Navier-Stokes equations, stationary vortices behind a cylinder are formed at
 moderate Reynolds number ($Re < 50$).  As the Reynolds number increases past $Re\approx50$, the configuration loses its symmetry and becomes unstable. 
New vortices
   then start to form alternately on both sides of the cylinder, while the
   vortices further downstream break away and develop into a Karman
   vortex street, as described by F\"oppl \cite{foeppl}.
    It has been argued by Roshko \cite{roshko} that ``possibly the breaking away should be regarded as primary,
resulting in asymmetry.'' The analysis presented in Sec.~\ref{sec:unst}
makes it clear that the reverse scenario is more plausible: the
asymmetrical disturbances induce the instability of the vortex pair
which then  breaks away from the cylinder. As vorticity is continuously
generated from the separated boundary layer on both sides of the
cylinder, new vortices are formed and alternately shed into the far wake
of the cylinder according to the unstable modes shown in
Fig.~\ref{fig:anti}.   Direct numerical simulations (DNS) of
two-dimensional flows past a cylinder performed by Tang and Aubry
\cite{tang} have confirmed that the mechanism for the instability of the
symmetric eddies in real flows is qualitatively described
by the instability of the point-vortex model.

It is experimentally observed \cite{roshko,roshko2} that vortex shedding
is suppressed if a splitter plate is installed behind the cylinder in
the center plane of the wake.  The presence of the splitter plate tends
to enforce symmetry of the flow with respect to the centerline, thus
effectively reducing the appearance of antisymmetric disturbances behind
the cylinder.  The suppression of vortex shedding in this case is thus
entirely consistent with the fact that the F\"oppl equilibria of the
vortex-point model is nonlinearly stable against symmetric perturbations and that
vortex shedding is induced by unstable antisymmetric modes, as discussed
above. This scenario has  been confirmed by  DNS of flows past a cylinder with symmetry imposed along the centerline recently
performed by Kumar {\it et al.} \cite{kumar}.  The problem of stationary configurations for vortex flows   past a cylinder with patches of constant vorticity has also been
  studied numerically by Elcrat {\it et al.}
  \cite{elcrat1,elcrat2}. These authors found two families of solutions, representing desingularized versions of the F\"oppl and the normal equilibria,  respectively, which have the same stability properties as the
 corresponding point-vortex equilibria. 

In conclusion, we have seen that the F\"oppl model, where a pair of counter-rotating point vortices move around a circular cylinder in the presence of a uniform stream, is a rich nonlinear dynamical system whose features---notably its stability properties---bear a direct relevance to our understanding of the vortex shedding mechanism in real flows. The results obtained here should, in principle, carry over to more general geometries, such as vortex motion around a plate or around a cylinder with noncircular cross section.

\begin{acknowledgments}
This work was supported in part by the Brazilian agencies CNPq and FACEPE.  One of the authors (AMJS) acknowledges financial support from CAPES, Brazil through a visiting professor scholarship. 
\end{acknowledgments}


\begin{thebibliography}{}
\bibitem{footnote} The formula for the matrix element $X'$ given in
  Eq.~(17) of Ref.~[5], which corresponds to the matrix element
  $B_{21}$  in  our Eq.~(\ref{eq:B21}), is in error.
\end{thebibliography}
\end{document}